\documentstyle[aps]{revtex} %galley

\input epsf

\preprint{WM-99-106; hep-ph/9903493}

\begin{document}
\twocolumn[
%-----------------------
\title{Soft Contributions to Hard Pion Photoproduction}
%-----------------------

\author{ Andrei Afanasev{\footnotemark}}
\address{ Department of Physics, North Carolina Central University,
Durham,  NC 27707\\ and 
Thomas Jefferson National Accelerator Facility,
12000 Jefferson Avenue, Newport News, VA 23606}

\author{ Carl E. Carlson and Christian Wahlquist}
\address{ Nuclear and Particle Theory Group, Physics Department, College of
William and Mary, Williamsburg, VA 23187-8795 }

%-------------------
\date{March 1999}
%------------------
\maketitle
%-------------------
%\begin{abstract}
\parshape=1 0.75in 5.5in \indent {\small {Hard, or high transverse
momentum, pion  photoproduction can be a tool for probing the
parton structure of the beam and target. We estimate the soft
contributions to this process, with an eye toward delineating the
region where perturbatively calculable processes dominate.  Our
soft process estimate is based on vector meson dominance and data
based parameterizations of semiexclusive hadronic cross sections. 
We find that soft processes dominate in single pion
photoproduction somewhat past 2 GeV transverse momentum at a few
times 10 GeV incoming energy.  The recent polarization asymmetry
data is consistent with the perturbative asymmetry being diluted by
polarization insensitive soft processes.  Determining the
polarized gluon distribution using hard pion photoproduction
appears feasible with a few hundred GeV incoming energy (in the
target rest frame). }
 
}

%\end{abstract}
%actual number now substituted for 9903493
\widetext 
\vglue -8.8cm \hfill JLAB-THY-99-07; WM-99-106; hep-ph/9903493
\vglue 8.8cm 
%-------------------------
%\pacs{13.88+e, 14.20.Gk, 13.60.Hb, 12.40.N, 12.38.Bx}

]

\narrowtext

%%%%%%%%%%%%%%%%%%%%%%%%%%%%%%%%%%%%%%%%%%%%%%%%%%%%%%%%%%%%%%%

\footnotetext{\vglue -25pt $^*$On leave from Kharkov
Institute  of Physics and Technology, Kharkov, Ukraine.}

%%%%%%%%%%%%%%%%%%%%%%%%%%%%%%%%%%%%%%%%%%%%%%%%%%%%%%%%%%%%%%%%%%%%%%%%%%

\section{Introduction}

%%%%%%%%%%%%%%%%%%%%%%%%%%%%%%%%%%%%%%%%%%%%%%%%%%%%%%%%%%%%

Recent results~\cite{anthony99} on pion photoproduction or, more
precisely, low-$Q^2$ electroproduction, show a need for a careful
estimate of soft contributions.  In particular, the measured
polarization dependent effects are not in good overall agreement
with calculations based only on QCD calculated using perturbation
theory (pQCD). A key issue is where and if the high transverse
momentum cross section is dominated by perturbatively calculable
contributions and where soft contributions are important.  In the
region where perturbative contributions dominate, it is known how
hard pion photoproduction can be a source of information about
hadron structure~\cite{acw97,acw98,many}.

Pion photoproduction at high transverse momentum, or
hard pion photoproduction, supplements what can be learned in the
standard hadron structure probes of deep inelastic scattering
and Drell-Yan processes, lately joined by high-$Q^2$ coincident
meson production~\cite{flavor}.  A particular feature of high
transverse momentum pion photoproduction with polarized initial
states is the sensitivity to the polarized gluon distribution,
$\Delta g$, in leading order.  This contrasts to the other
processes mentioned, which have no leading order gluon
contribution.  Additionally, in some kinematic regions the process
occurs mainly due to pion production at short distances (``direct
pion production''), whereupon there is sensitivity to the high-$x$
valence quark distribution and the short distance pion wave
function.

Many authors have calculated perturbative contributions to hard pion
photoproduction, and recent work in this area has centered on
short distance pion production~\cite{acw97},
polarization effects~\cite{acw98,many}, and complete next to leading
order corrections~\cite{many}. These calculations do include the
hard or short distance contributions from hadronic components
of the photon, under the heading of resolved photon processes, but
do not include the soft part. Here we present a phenomenological
calculation of soft contributions, and compare its size to the pQCD
results already known.  The calculation relies on vector dominance
(VMD), which is a way to represent the hadronic components of the
photon as they enter into soft processes.  Experimental studies, in
particular the Omega~\cite{omega}, H1~\cite{h1}, and
Zeus~\cite{zeus} collaborations, have shown that hadron induced and
photon induced hadron production were proportional to each other up
to a certain transverse momentum, and that above this transverse
momentum the photon induced reactions rise relative to hadron
induced ones as the pointlike piece of the photon becomes more
important.  For the kinematics of the above experiments it is about
2 GeV transverse momentum where the pointlike photon
begins to become apparent.

In the next section, we put together known photon vector meson
couplings with phenomenological representations of the hadron-hadron
reactions to produce soft cross section formal results for
kinematics of interest.  Following that, section~\ref{results}
presents some numerical results for cross sections and polarization
effects, making an assumption that the polarization dependence of
the soft processes is small.  We close with a discussion in
section~\ref{discussion}.

%%%%%%%%%%%%%%%%%%%%%%%%%%%%%%%%%%%%%%%%%%%

\section{Outline of Calculations}     \label{calc}

%%%%%%%%%%%%%%%%%%%%%%%%%%%%%%%%%%%%%%%%%%%

We are aware that quite successful descriptions of soft
processes have been obtained using Regge theory inspired
models~\cite{torbjorn}.  However, the sophistication of these
models makes them somewhat rich in parameters that need to be
set from the same data that is being described, or from similar
data.  For example, there is a need for some cutoffs whose scale
parameters are not predicted from theory, a use of different
Pomeron intercepts for single diffractive processes and total
cross sections, and a fitting of the overall size in the form
of the triple Pomeron coupling using related reactions. We opt
for a complementary course, wherein we simply use known
couplings to calculate photon to vector meson conversion and
then use measured data for the hadronic cross sections.

For definiteness we will consider $\pi^+$ production off a
proton target.

The $\rho$-dominance amplitude is

\begin{equation} 
\left. f(\gamma p \rightarrow \pi^+ X) \right|_\rho
  = {e\over f_\rho} f(\rho^0 p \rightarrow \pi^+ X)
\end{equation}

\noindent and

\begin{eqnarray}
 &d\sigma& (\gamma p \rightarrow \pi^+ X) 
  = {\alpha\over \alpha_\rho} 
    d\sigma (\rho^0 p \rightarrow \pi^+ X)
                                          \nonumber \\
    &+& {\rm other\ VMD} + {\rm non\ VMD\ contributions}  ,
\end{eqnarray}

\noindent where $\alpha_\rho \equiv f_\rho^2/4\pi$ and `other
VMD' stands for contributions of excited $\rho$'s and of
other vector mesons.  The value of
$\alpha_\rho$ can be got from 
$\Gamma(\rho \rightarrow e^+ e^-)$ and is~\cite{ps97}
\begin{equation}
\alpha_\rho = 2.01 \pm 0.10 .
\end{equation}

This reduces the problem to finding the cross section or a
parameterization thereof for vector meson production of the
$\pi^+$. In principle, this might be an experimentally
measurable process, but in practice we will have to approximate
it be charged pion induced processes.  The remainder of this
section is mostly devoted to explaining how we do this.  First
we make some remarks on contributions from excited $\rho$'s
and other vector mesons.

Excited $\rho$ contributions to 
$\gamma p \rightarrow \rho + X$  decrease the rate by
20\%,  according to Pautz and Shaw~\cite{ps97}.  The basic relation is
\begin{equation}
f(\gamma p \rightarrow \rho X)
  = {e\over f_\rho} 
   \underbrace{f(\rho p \rightarrow \rho X)}
        _{\displaystyle{``f_{\rho\rho}"}}  
  + {e\over f_{\rho'}} 
   \underbrace{f(\rho' p \rightarrow \rho X)}
        _{\displaystyle{``f_{\rho'\rho}"}}
\end{equation}
and the claim is that while the couplings are about the
same, the amplitudes interfere destructively,
\begin{equation}
f_{\rho'\rho} \approx (-16\%) f_{\rho\rho}  .
\end{equation}
The effect can be subsumed by simply calculating
simple vector meson dominance with 
$\alpha_\rho^{eff} = 2.44$.  For us the question is whether the same
is true for $\pi^+$ production,

\begin{equation}
f_{\rho'\pi} \stackrel{?}{\approx} (-16\%) f_{\rho\pi}  ,
\end{equation}

\noindent and we shall proceed assuming it is true.

From flavor SU(3), the couplings of the photon to the
vector mesons lie in the ratios
\begin{equation}
f_\rho^{-2} : f_\omega^{-2}  :  f_\phi^{-2} 
  =  9 : 1 : 2  .
\end{equation}

\noindent {\it If} the $\rho$, $\omega$, and $\phi$ strong
interaction cross sections are the same, then the other flavors
add 33\% to the $\rho$ contribution.  At the present level of
knowledge, we will approximate the total VMD contribution by the
$\rho$ contribution multiplied by 4/3.
The photoproduction cross section is now

\begin{eqnarray}
d\sigma (\gamma p &\rightarrow& \pi^+ X)
  = {4\over 3} {\alpha\over \alpha_\rho^{eff}} 
    d\sigma (\rho^0 p \rightarrow \pi^+ X)
                                        \nonumber \\
   &+& {\rm non\ VMD\ contributions}  ,
\end{eqnarray}

\noindent with $\alpha_\rho^{eff} = 2.44$.  Off shell effects have
not been considered.

We need knowledge, or a representation, of 
$d\sigma(\rho^0 p\rightarrow \pi^+ X)$.  Often used is,
\begin{eqnarray}
d\sigma(\rho^0 p \rightarrow \pi^+ X)
 &=& {\scriptstyle {1 \over 2}}
   d\sigma(\pi^+ p \rightarrow \pi^+ X)
                                         \nonumber \\
 &+& {\scriptstyle {1 \over 2}}
   d\sigma(\pi^- p \rightarrow \pi^+ X)  .
\end{eqnarray}
This will not work in the forward direction, where one
cross section has a leading particle effect but
$\rho^0 p \rightarrow \pi^+ X$ should not.  One may expect the
measurable cross section most similar to 
$\rho^0 p \rightarrow \pi^+ X$ would be 
$\pi^+ p \rightarrow \pi^0 X$.  
Data from O'Neill {\it et al.}~\cite{oneill76} show that 
$\pi^+ p \rightarrow \pi^0 X$ has the same angular
dependence as $\pi^+ p \rightarrow \pi^- X$ but is about
30\% larger.  This reduces the problem to finding a representation
of the latter.

Bosetti {\it et al.}~\cite{bosetti73}, who
experimentally studied charged pion cross sections, found that the
cross section $\pi^+ p \rightarrow \pi^- X$ factors in $k_T$ and
$\xi$, where $\xi$ is the scaled rapidity,
\begin{equation}
\xi = {y - y_t \over y_p - y_t}
\end{equation}
for $p=$ projectile and $t =$ target, and $y$ is the
rapidity, which may be defined in various equivalent ways including
\begin{equation}
y = {\rm arcsinh} {p_L \over \sqrt{p_T^2 + m^2}}.
\end{equation}
That means, 
\begin{equation}
\omega_\pi {d\sigma \over d^3k}
  = \left. \omega_\pi {d\sigma \over d^3k} \right|_
        {90^\circ {\rm CM}}
 \times g(\xi) ,
\end{equation}
where $g(\xi)$ will have some dependence on $k_T$ to
respect kinematic bounds.  A choice that appears to work
is
\begin{equation}
g(\xi) = \left( 1 -
     {(\xi - \xi_0)^2 \over (\xi_{max} - \xi_0)^2 }
    \right)^2  ,
\end{equation}
where $\xi_0$ is is halfway between $\xi_{max}$ and $\xi_{min}$, and
$\xi_{max,min}$ are the maximum or minimum $\xi$ for a given $k_T$.

Beier {\it et al.}~\cite{beier78} have analytic forms that work over a
wide kinematic range for 
$pp \rightarrow \pi^- X$ at 90$^\circ$ in the CM, and the Bosetti
{\it et al.} data~\cite{bosetti73} approximately agree with
\begin{equation}
\omega_\pi {d\sigma \over d^3k}
      (\pi^+ p \rightarrow \pi^- X) = {2 \over 3}
\omega_\pi {d\sigma \over d^3k}
      (pp \rightarrow \pi^- X)  .
\end{equation}

In summary, calculate using
\begin{eqnarray}
&\omega_\pi& {d\sigma \over d^3k}
     (\gamma p \rightarrow \pi^+ X)
  = 
                                    \nonumber \\
     &+&    {\alpha \over \alpha_\rho^{eff}}
      \cdot 1.3 \cdot {4\over 3} \cdot {2\over 3} \cdot 
    \left. \omega_\pi {d\sigma \over d^3k}
      (pp \rightarrow \pi^- X) \right|_{90^\circ {\rm CM}}
     g(\xi)  
                \nonumber \\[1.5ex]
  &\qquad& +  {\rm \ non\ VMD\ contributions} .
\end{eqnarray}

The ``non VMD'' contributions are discussed in, for
example,~\cite{acw97,acw98,many}. To review the numerical factors,
the single charge change reaction was about 1.3 times the double
charge change reaction according to~\cite{oneill76}, the 4/3 is to
account for the $\omega$ and $\phi$ mesons, and the 2/3 is so that
we may use $pp$ cross section parameterizations as stand-ins for
meson-proton cross sections.  Electroproduction data with particle
identification with electron energies up to 19 GeV, reported in
Wiser's thesis~\cite{wiser}, indicate that $\pi^-$ production off a
proton target is about a factor 1.3 lower than for the $\pi^+$, and
that $\pi^\pm$ is produced off a neutron at about the same rate as
$\pi^\mp$ off a proton.

The connection between photoproduction and electroproduction when the
outgoing electron is unobserved is given by the Weiz\"acker-Williams
equivalent photon approximation,

\begin{equation}
d\sigma(eN\rightarrow \pi X) = \int^{E_e}_{E_{min}}
    dE_\gamma \, N(E_\gamma) d\sigma(\gamma N \rightarrow \pi X) .
\end{equation}

\noindent The expressions we use for the photon number density
$N(E_\gamma)$ and the lower limit are quoted in~\cite{acw98}.

%%%%%%%%%%%%%%%%%%%%%%%%%%%%%%%%%%%%%%

\section{Some Results}     \label{results}

%%%%%%%%%%%%%%%%%%%%%%%%%%%%%%%%%%%%%%

\begin{figure}  

\epsfxsize 3.3 in
\centerline{  \epsfbox{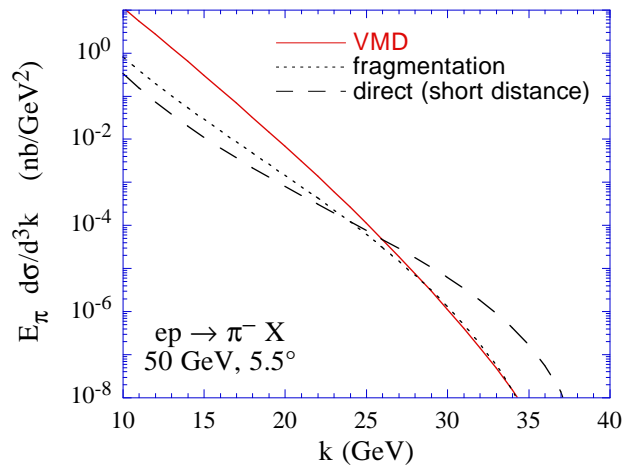}  }

\vglue 1cm

\epsfxsize 3.3 in
\centerline{  \epsfbox{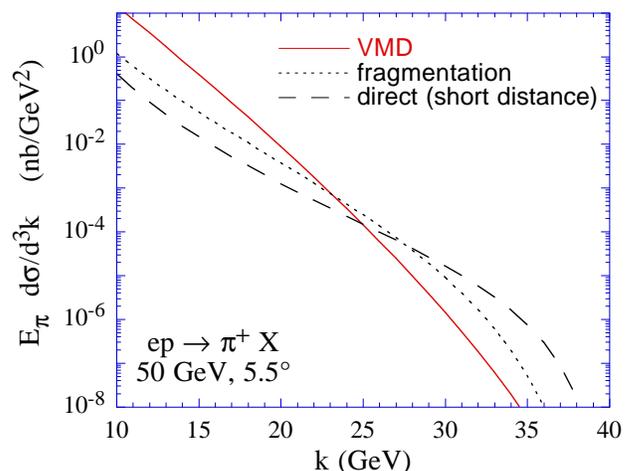}   }

\caption{The invariant differential cross sections for 
$ep \rightarrow \pi^- X$, above, and $\pi^+$, below.  The incoming
electron energy is 50 GeV, and pion lab angle is 5.5$^\circ$.}

\label{xsection}

\end{figure}

We begin by examining the differential cross section for one relevant
kinematic situation, namely that with 50 GeV incoming electrons with
pions emerging at 5.5$^\circ$ in the lab. This energy is typical of
SLAC and not far above what can be obtained at HERMES.
Fig.~\ref{xsection} shows the unpolarized differential cross
section vs. pion momentum for both the $\pi^-$ and $\pi^+$.  There
are three curves on each plot, the soft contribution represented by
VMD and two perturbative contributions, namely parton production
followed by fragmentation and direct or short range pion
production.  (Another perturbative contribution, the resolved
photon process is small enough at this energy and angle not to be
an issue.)  The three contributions should be added incoherently.

The soft contribution continues to a momentum that is higher than
expected.  Nonetheless, one sees that at momenta beyond about 25 GeV
for the $\pi^-$ or 22 GeV for the $\pi^+$,  the sum of the
perturbative contributions exceed the soft contributions.  For this
angle, this is about 2.4 and 2.1 GeV of transverse momentum,
respectively.

%%%%%%%%%%%%%%%

%%%%%%%%%%%%%%%%

\begin{figure}

\epsfxsize 3.3 in
\centerline{  \epsfbox{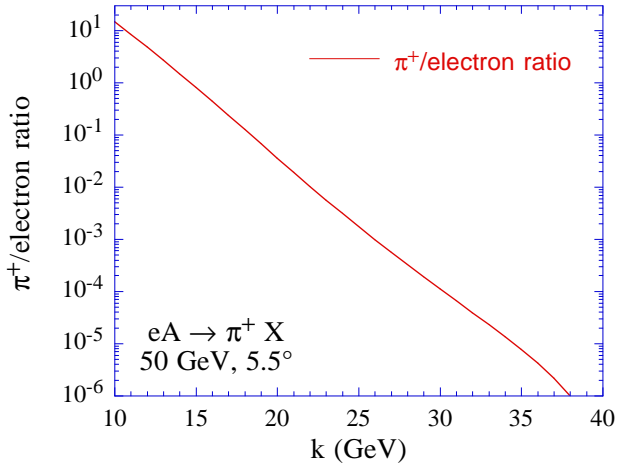}   }

\caption{The calculated pion to electron ratio for a 
$^{15}$N He$_3$ target. For this target and this vertical
scale, the $\pi^-$ results are hardly distinct from the
$\pi^+$. They are in reasonable accord with data as reported
in~\protect\cite{anthony99}. }

\label{pieratio}

\end{figure}

The hadron to electron ratio is also measured and reported in the
experimental paper~\cite{anthony99}.  At lower momenta the
calculated $\pi$/e ratio is too small without the VMD
contributions.  With all contributions added together, the
calculated pion to electron ratio is shown in Fig.~\ref{pieratio}. 
These are in reasonable accord with the plots presented
in~\cite{anthony99}, which in turn are stated to be in reasonable
accord with the data.

Having a reasonable description of the unpolarized cross section in
hand, we need to consider the polarization asymmetry.  If $R$ and
$L$ represent photon helicities and $\pm$ represent target
helicities, then the longitudinal asymmetry $E$ or $A_{LL}$ is
defined by

\begin{equation}
{E} = A_{LL} \equiv {\sigma_{R+} - \sigma_{R-}
   \over
   \sigma_{R+} + \sigma_{R-} }.
\label{asym}
\end{equation}

\noindent The
polarization dependence of the perturbative terms is calculable, but
we have no direct polarization information on the VMD
contributions.  One class of VMD subprocess would give a negative
polarization asymmetry if hadron helicity conservation holds for
those diagrams.  This is the reaction
$V + q \rightarrow \pi + q$, where $V$ stands a vector meson which
must have helicity $\pm 1$ since it comes from conversion of a real
photon.  The the vector meson and initial quark must have opposite
helicity, or else the final state must have total helicity $\pm
1/2$.  However, Manayenkov~\cite{m99} has argued from a Regge
analysis that the soft contributions to $A_{LL}$ are small.  Here,
we shall assume no polarization dependence for the VMD terms.  The
polarization asymmetry then comes only from the perturbative terms,
but it is much muted at low momentum because of the large
non-perturbative cross section.

%%%%%%%%%%%%%%%

%%%%%%%%%%%%%%%%

\begin{figure}  

\epsfxsize 3.3 in
\centerline{  \epsfbox{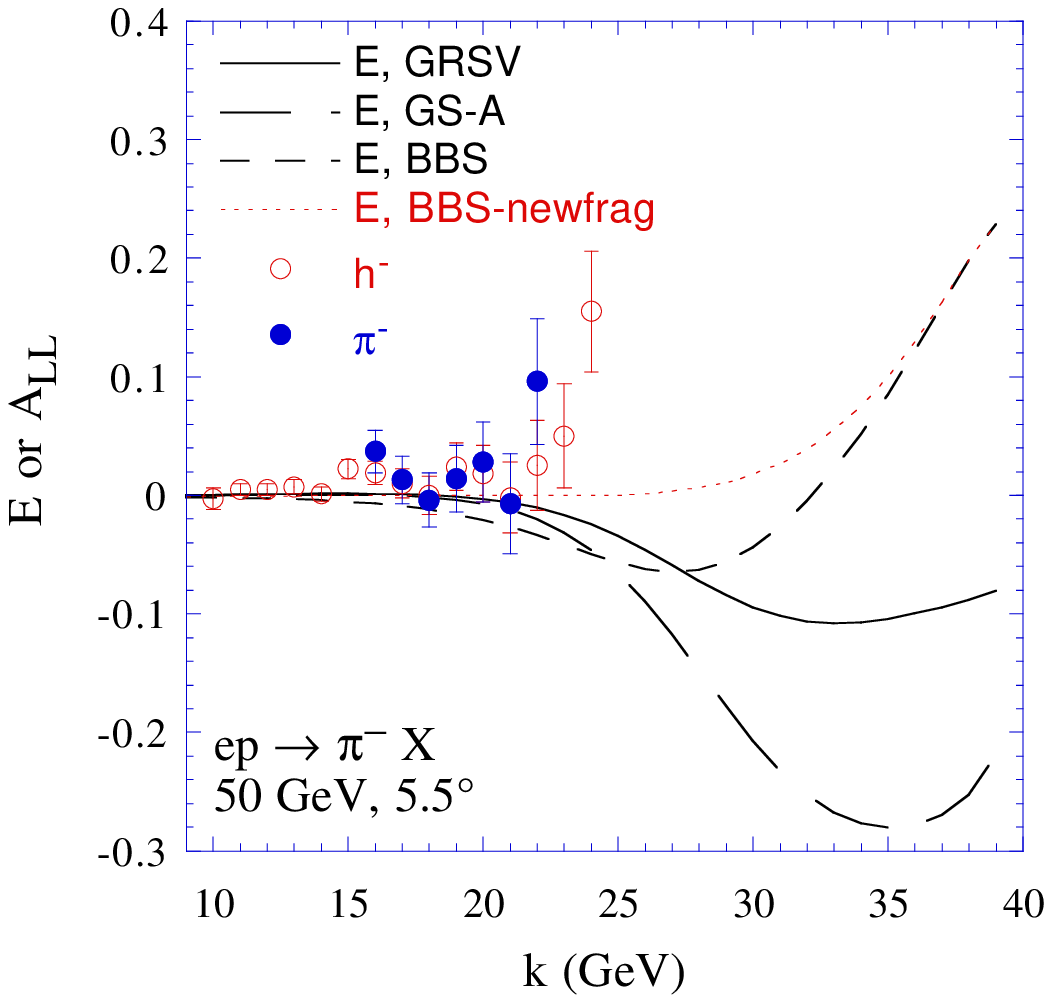}  }

\epsfxsize 3.3 in
\centerline{  \epsfbox{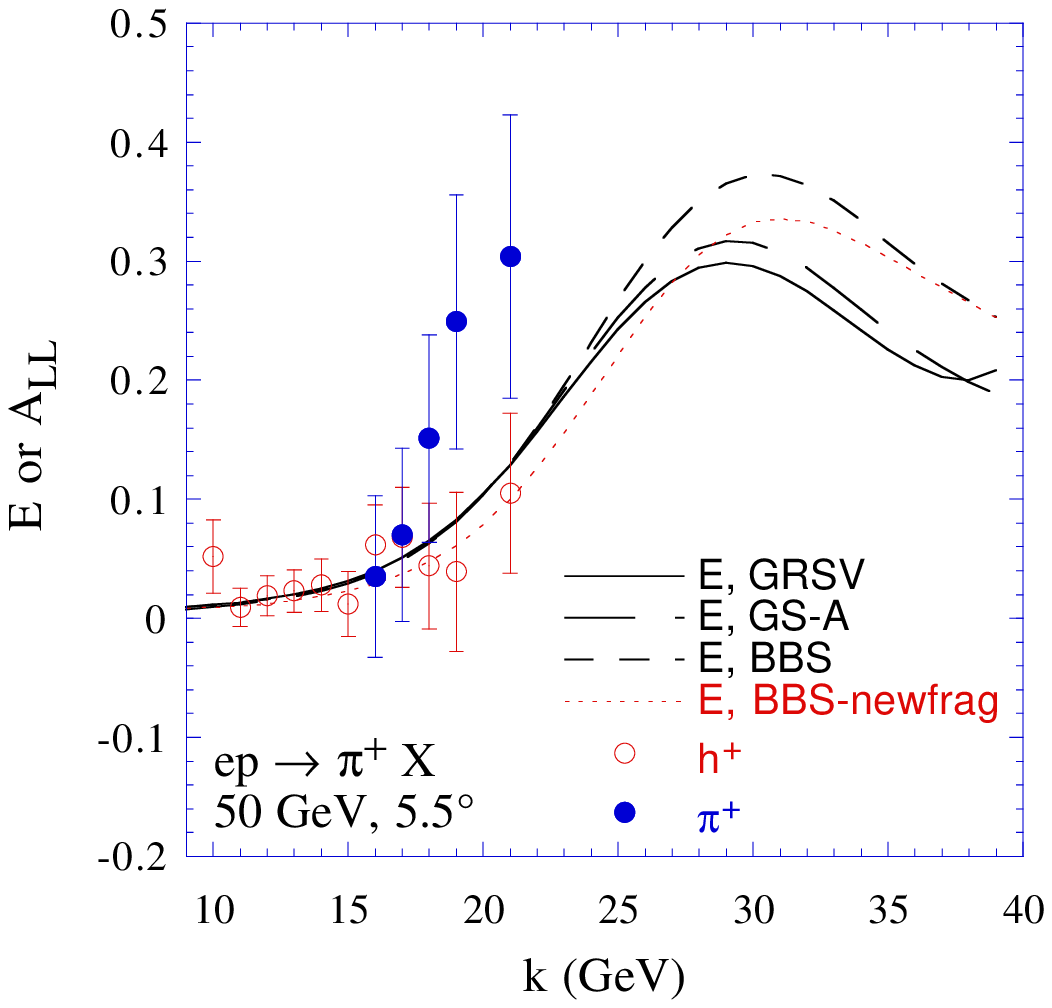}   }

\caption{Polarization asymmetries for $ep \rightarrow \pi^-$, above
and $\pi^+$, below.  The data for charged hadrons and for charged
pions is from~\protect\cite{anthony99}.}

\label{polarization1}

\end{figure}

\begin{figure}  

\epsfxsize 3.3 in
\centerline{  \epsfbox{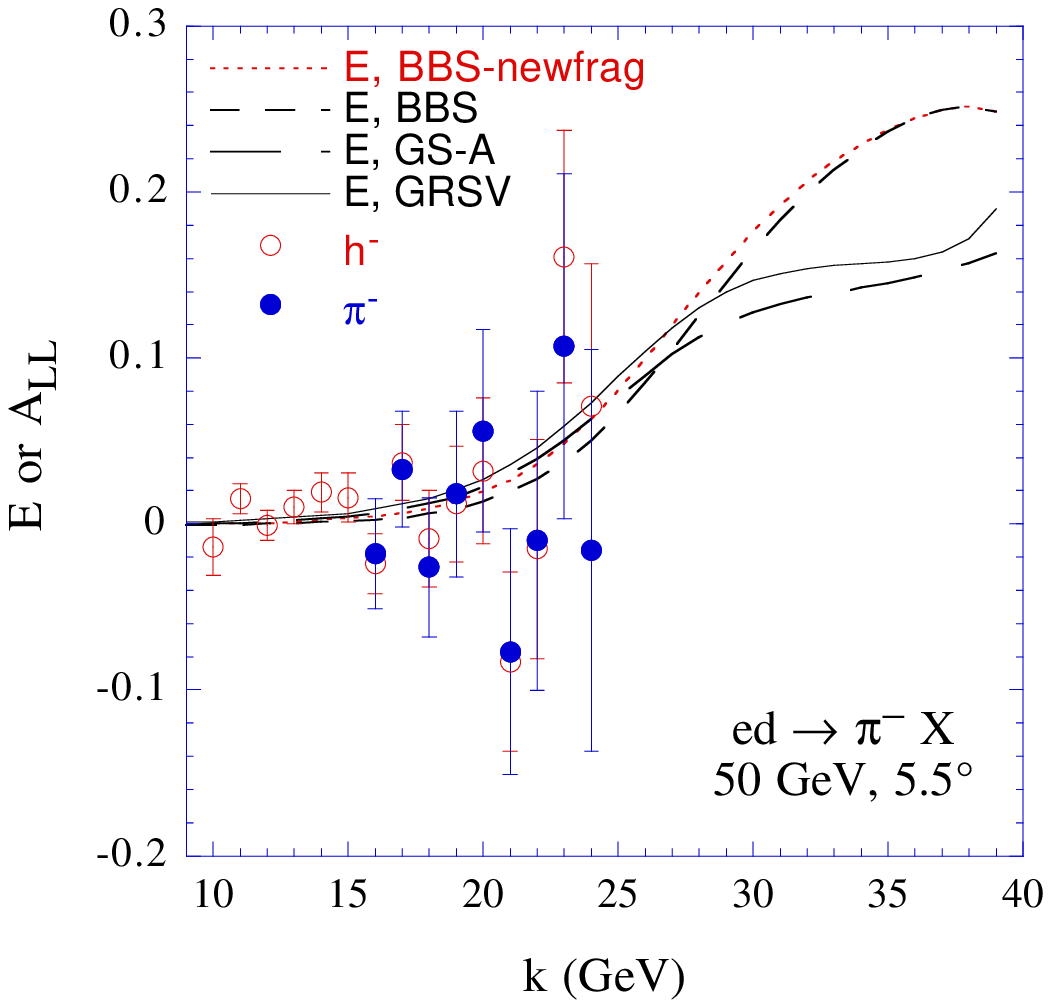}  }

\epsfxsize 3.3 in
\centerline{  \epsfbox{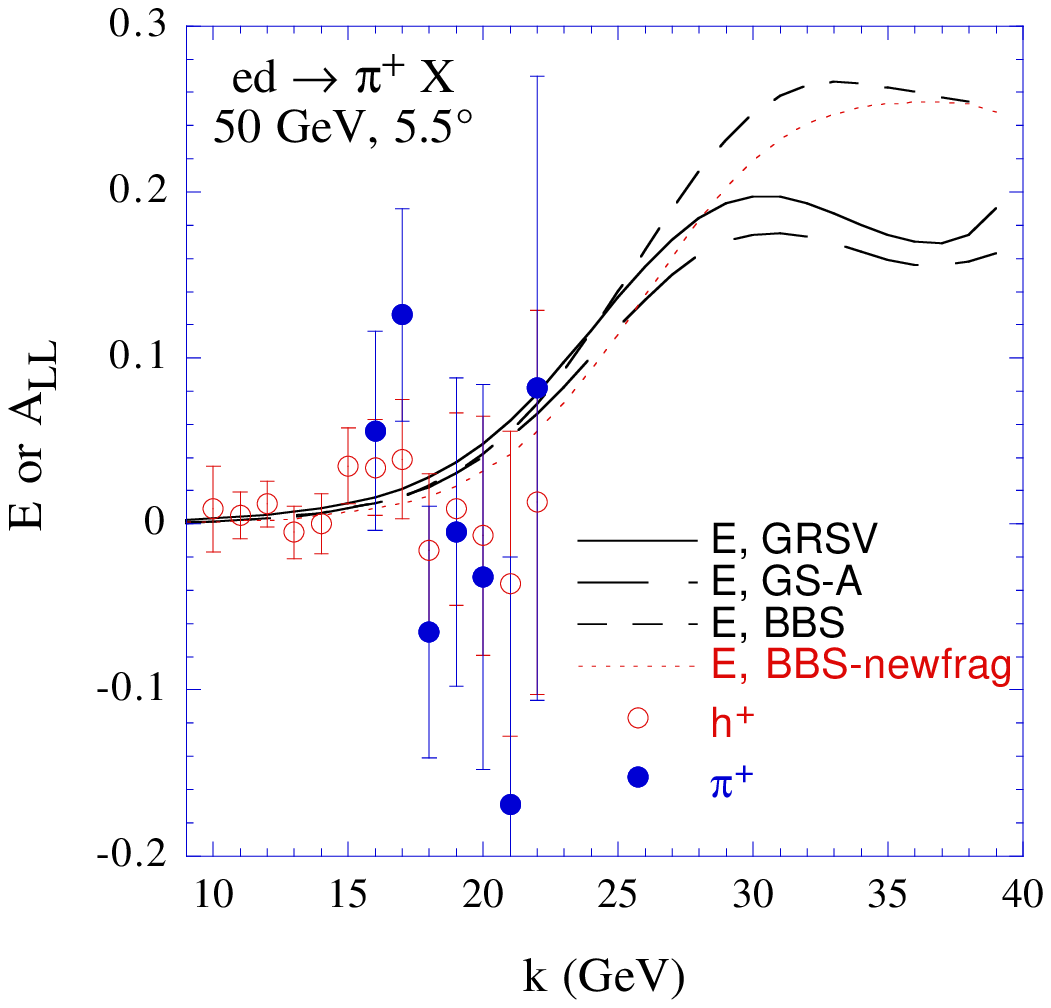}   }

\caption{Polarization asymmetries for the deuteron, with 
$ed \rightarrow \pi^- X$ above and $ed \rightarrow \pi^+ X$ below. 
The data is from~\protect\cite{anthony99}.}

\label{polarization2}

\end{figure}

Actual polarization asymmetry results plotted vs. pion momentum,
again for electron energy 50 GeV and pion angle 5.5$^\circ$, are shown
for proton targets in Fig.~\ref{polarization1} and for deuteron
targets in Fig.~\ref{polarization2}.   (The  deuteron in this
calculation is treated simply as a proton plus a neutron.) Also shown
are polarization asymmetry data~\cite{anthony99} for charged hadrons
and for identified $\pi^\pm$.

A few words should be said about the distribution functions and
fragmentation functions.  GRSV~\cite{grsv96} and GS~\cite{gs96} are
both widely used, and BBS~\cite{bbs95} differ from them most notably
in having the pQCD counting rule results for the d-quark to u-quark
ratio for large
$x$, and by not nicely separating sea quark contributions.  Since for
us the distribution functions are most needed at large $x$, the latter
may not be so serious.  The d/u ratio now appears, with more careful
examination of how the neutron structure functions are extracted from
deuteron data~\cite{wally}, to the pQCD ratio of 1/5 rather than to
zero, which makes it important to notice how different the BBS
results are from the others at high momenta.

We used the fragmentation functions given in~\cite{cw93} and our
experience has been that the results at least at SLAC or HERMES
energies would not be too different for the $\pi^+$ but larger in
magnitude for the $\pi^-$ if we used~\cite{bkk95}.  However, recent
HERMES data suggests that the ``unfavored'' fragmentation function
(e.g., for a u-quark fragmenting to a $\pi^-$) is larger than what
we have been using~\cite{makins}.  So for the BBS distribution, we
present results from one additional fragmentation function, where the
sum determined from
$e^+ e^- \rightarrow \pi X$ is unchanged but the ratio of unfavored or
secondary fragmentation function to primary (or favored minus
unfavored) fragmentation function is given by

\begin{equation} 
D_s(z)/D_p(z) = 0.5 (1-z)^{0.3}/z  ,
\end{equation}

\noindent where $z$ is the fraction of quark momentum that goes into
the pion.

%%%%%%%%%%%%%%%

%%%%%%%%%%%%%%%%

%%%%%%%%%%%%%%%

%%%%%%%%%%%%%%%%

%%%%%%%%%%%%%%%%%%%%%%%%%%%%%%%%%%%%%%%%%%%%%%%%%%%%%%

\section{Discussion}            \label{discussion}

%%%%%%%%%%%%%%%%%%%%%%%%%%%%%%%%%%%%%%%%%%%%%%%%%%%%%%

We believe we have presented as accurate an estimate of the soft
processes in pion photoproduction as can currently be done. 
Improvements could follow given more information.  
For examples, the connections we made in section~\ref{calc} require
some leaping among processes, and we have not included pions from
target fracture in the perturbative cases, nor have we deeply
entered into the questions newly revived about the unfavored
fragmentation functions.  We feel the latter is an important
question that should be the subject of a separate study.  Having
made our caveats, we do have a clear and logical representation of
the soft contributions that we can compare to the newest pion
photoproduction (or low $Q^2$ electroproduction) data.  We find
that the soft process, working through VMD, can explain the total
cross section at lower transverse momentum.

We find further that the data is compatible with the idea that there
is little polarization asymmetry in the soft interactions, as may be
seen in our comparisons to the data in Figs.~\ref{polarization1}
and~\ref{polarization2}.  We would like to be able to confirm or
understand this by other means.

\begin{figure}

\epsfxsize 3.3 in
\centerline{  \epsfbox{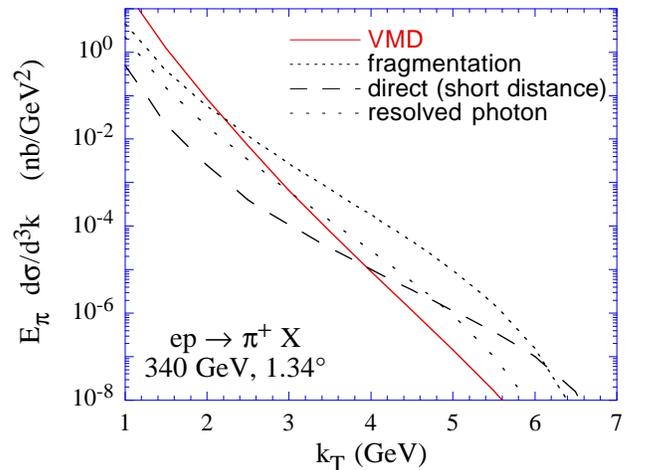}   }

\caption{The differential cross section
for 340 GeV electrons impinging on an standing proton with positive
pions emerging at 1.34$^\circ$.  (This corresponds to a collider with
4 GeV electrons hitting 40 GeV protons and pions emerging at
90$^\circ$ in the lab.)}

\label{xsection340}

\end{figure}

Perturbation theory can be used to calculate the
cross section and polarization dependence at higher transverse
momentum.  The crossover is at a bit over 2 GeV for the kinematic
regions we have dealt with here.  The idea that hard pion
photoproduction is sensitive to $\Delta g$ is true in a region where
pQCD is valid and the fragmentation process dominates. As a reminder,
it is true because the gamma-gluon fusion process accounts for a
reasonable fraction of the hard pion photoproduction, and this
process has a magnitude 100\% polarization asymmetry. However, it
requires somewhat more energy so that there is a region above the VMD
region where the fragmentation process is important. As an example, we
present in Fig.~\ref{xsection340} a differential cross section
for 340 GeV electrons impinging on an standing proton with pions
emerging at 1.34$^\circ$.  (This corresponds to a collider with 4 GeV
electrons hitting 40 GeV protons and pions emerging at 90$^\circ$ in
the lab. The energies are pertinent to an Electron
Polarized Ion Collider under discussion at the Indiana University
Cyclotron Facility.)  We see the sort of region we want between
about 2 and 6 GeV of transverse momentum.

%%%%%%%%%%%%%%%

%%%%%%%%%%%%%%%%

We have been greatly motivated by the idea that hard pion
photoproduction can give information on parton distributions.  We
note that this is already proving feasible.  The H1 collaboration,
working in a region where the resolved  photon process dominates, has
extracted the gluon density in the photon from data on this
process~\cite{h1_adloff}.

The idea that the ratio $d(x)/u(x)$ obeys the pQCD limit for large
$x$, rather than falling to zero, is gaining ground.  So far the
relevant analyses~\cite{wally} are only for the unpolarized case, but
the $x \rightarrow 1$ polarization prediction of 100\% polarization
parallel to the parent hadron can be tested here.  With direct pion
production (or also with fragmentation) off valence quarks dominant
at the highest
$E_\pi$, one has an asymmetry for the $\pi^-$ of

\begin{equation}
E = A_{LL} = {s^2-u^2 \over s^2+u^2} \cdot {\Delta d \over d}
           = 0.24 {\Delta d \over d}
\end{equation}

\noindent where the number is for $E_e = 50$ GeV, pion
$\theta_{lab}=5.5^\circ$, and the highest allowed $E_\pi$ (in this
case, 41.2 GeV).  For pQCD as $x \rightarrow 1$ one has 
$\Delta d = d$, and one can see this trend in the results for the
BBS~\cite{bbs95} distribution functions since BBS follows the pQCD
limit.  In fact, for pQCD the limiting asymmetry is the same for
$\pi^\pm$ and independent of target.

\section*{acknowledgments}

We thank Peter Bosted and the E155 collaboration for discussion of
their data, and thank Torbj\"orn Sj\"ostrand for a helpful
communication.  AA thanks the DOE for support under grant
DE-AC05-84ER40150; CEC and CW thank the NSF for support under
grant PHY-9600415.


\begin{thebibliography}{99}

\bibitem{anthony99} P. L. Anthony {\it et al.}, SLAC-PUB-8049 (1999).

\bibitem{acw97} A. Afanasev, C. Carlson, and C. Wahlquist,
                Phys. Lett. B {\bf 398}, 393 (1997).

\bibitem{acw98} A. Afanasev, C. Carlson, and C. Wahlquist,
                Phys. Rev. D {\bf 58}, 054007 (1998).

\bibitem{many} D. De Florian and W. Vogelsang,
               Phys. Rev. D {\bf 57}, 4376 (1998);
               B. A. Kniehl, Talk at Ringberg Workshop,
               hep-ph/9709261; 
               M. Stratmann and W. Vogelsang,
               Talk at Ringberg Workshop, hep-ph/9708243.

\bibitem{flavor} L. L. Frankfurt {\it et al.},
              Phys. Lett. B {\bf 230}, 141 (1989);
              F. E. Close and R. G. Milner,
              Phys. Rev. D {\bf 44}, 3691 (1991);
              B. Adeva {\it et al.}, 
              Phys. Lett. B {\bf 369}, 96 (1996).

\bibitem{omega} R. J. Apsimon {\it et al.}, 
                Z. Phys. C {\bf 43}, 63 (1989).

\bibitem{h1} I. Abt {\it et al.}, 
             Phys. Lett. B {\bf 328}, 176 (1994).

\bibitem{zeus} M Derrick {it et al.}, 
               Z. Phys. C {\bf 67}, 227 (1995).

\bibitem{torbjorn} G. Schuler and T. Sj\"ostrand,
                   Nucl. Phys. B {\bf 407}, 539 (1993);
                   T. Sj\"ostrand, 
                   J. Phys. G {\bf 22}, 709 (1996).
                   The physics described in these articles has
                   been implemented as part of the PYTHIA event
                   generation code.  See T. Sj\"ostrand, 
                   Comput. Phys. Commun. {\bf 82}, 74 (1994).

\bibitem{ps97} A. Pautz and G. Shaw, hep-ph/9710235.

\bibitem{oneill76} L. H. O'Neill {\it et al.}, 
                   Phys. Rev. D {\bf 14}, 2878 (1976).

\bibitem{bosetti73} P. Bosetti {\it et al.},
                    Nucl. Phys. B {\bf 54}, 141 (1973).

\bibitem{beier78} E. Beier {\it et al.},
                  Phys. Rev. D {\bf 18}, 2235 (1978).

\bibitem{wiser} D. Wiser, Ph. D. thesis, 
                University of Wisconsin, 1977 (unpublished).

\bibitem{m99} S. I. Manayenkov, report DESY 99-016, hep-ph/9903405.
               

\bibitem{grsv96} M. Gl\"uck, E. Reya, M. Stratmann,
                 and W. Vogelsang,
                 Phys. Rev. D {\bf 53} 4775 (1996).

\bibitem{gs96} T. Gehrmann and W. J. Stirling,
               Phys. Rev. D {\bf 53}, 6100 (1996).

\bibitem{bbs95} S. J. Brodsky, M. Burkardt, and I. Schmidt, 
              Nucl. Phys. B {\bf 441}, 197 (1995).

\bibitem{wally}  W. Melnitchouk, J. Speth, and A. W. Thomas,
                 Phys. Lett. B {\bf 435}, 420 (1998);
                 W. Melnitchouk and J.C. Peng,
                 Phys. Lett. B {\bf 400}, 220 (1997);
                 W. Melnitchouk and A. W. Thomas,
                 Phys. Lett. B {\bf 377}, 11 (1996);
                 U. K. Yang and A. Bodek, hep-ph/9809480.

\bibitem{cw93} C. E. Carlson and A. B. Wakely,
               Phys. Rev. D {\bf 48},  2000 (1993).

\bibitem{bkk95} J. Binneweis, B. A. Kniehl, and G. Kramer,
                Z. Phys. C {\bf 65}, 471 (1995) and 
                Phys. Rev. D {\bf 52}, 4947 (1995).

\bibitem{makins} N. Makins, Proceedings of CEBAF Workshop on
                 Physics and Instrumentation with 6-12 GeV Beams,
                 ed. S. Dytman, H. Fenker, and P. Roos,
                 JLab, Newport News, June 1998, p.97;
         Ph. Geiger, {\it Measurement of Fragmentation Functions at
         HERMES}, Ph. D. Thesis, Ruprechet-Karls-Universit\"at,
         Heidelberg, 1998.

\bibitem{h1_adloff} C. Adloff {\it et al.}, hep-ex/9810020.

\end{thebibliography}
\end{document}